\documentclass[amsfonts,amsmath,amssymb,letterpaper,twocolumn,pra,superscriptaddress]{revtex4}

\usepackage{ifpdf}
\ifpdf
  \usepackage[pdftex]{graphicx}
  \usepackage{epstopdf}
\else
  \usepackage[dvips]{graphicx}
\fi

\usepackage{psfrag}

\usepackage{amsfonts}
\usepackage{amssymb}
\usepackage{amsmath}
\usepackage{braket}
\usepackage{bm} %Provides bold Greek and math smbols via the \bm macro
\usepackage{hyperref}
\usepackage{xspace} % For sapcing after \ensuremath{}

\newcommand{\vectornorm}[1]{{\ensuremath{\left| #1 \right|}}}

\DeclareMathOperator*{\Tr}{Tr}

\newcommand{\Eref}[1]{Eq.~\eqref{#1}}

\newcommand{\Fref}[1]{Fig.~\ref{#1}}

\newcommand{\Sref}[1]{Sec.~\ref{#1}}

\newcommand{\Aref}[1]{Appendix~\ref{#1}}

\newcommand{\Z}{Z}

\newcommand{\X}{X}

\newcommand{\Y}{Y}
\newcommand{\I}{I}

\newcommand{\neigh}{\mathcal{N}}

\newcommand{\gen}{K}

\newcommand{\Mu}{{\ensuremath{\bm{\mu}}}}
\newcommand{\Nu}{{\ensuremath{\bm{\nu}}}}

\newcommand{\A}{\ensuremath{A}\xspace}
\newcommand{\B}{\ensuremath{B}\xspace}
\newcommand{\zero}{{\ensuremath{\rho^{(0)}}}\xspace}
\newcommand{\one}{{\ensuremath{\rho^{(1)}}}\xspace}
\newcommand{\two}{{\ensuremath{\rho^{(2)}}}\xspace}

\def\expec#1{\ensuremath{\mathinner{\langle#1\rangle}}\xspace}
\def\avg#1{\expec{\gen_{#1}}}
\def\avgj{\avg{j}}
\renewcommand{\vec}[1]{\ensuremath{\bm{#1}}}
\DeclareMathOperator{\T}{\ensuremath{T}\xspace}

\newcommand{\R}{R}

\newcommand{\bin}{\ensuremath{\lbrace0,1\rbrace}}

\newcommand{\stab}[1]{\gen_{\vec{#1}}}

\newcommand{\astab}[1]{\avg{\vec{#1}}}

\newcommand{\kket}[1]{\gen_{#1}}

\newcommand{\kron}[2]{\delta _{\vec{#1},\vec{#2}}}

\newcommand{\meas}[1]{\lambda^{(#1)}}
\newcommand{\measv}[1]{\vec{\lambda}^{(#1)}}

\newcommand{\D}{{\ensuremath{{\cal{D}}}}\xspace}

\newcommand{\Err}{{\ensuremath{{\cal{E}}}}\xspace}

\newcommand{\ab}{\vec{a}}
\newcommand{\bb}{\vec{b}}
\newcommand{\azero}{\ensuremath{\ab^{(0)}}\xspace}
\newcommand{\bzero}{\ensuremath{\bb^{(0)}}\xspace}
\newcommand{\aone}{\ensuremath{\ab^{(1)}}\xspace}
\newcommand{\bone}{\ensuremath{\bb^{(1)}}\xspace}
\newcommand{\atwo}{\ensuremath{\ab^{(2)}}\xspace}
\newcommand{\btwo}{\ensuremath{\bb^{(2)}}\xspace}

\begin{document}

\title{Purification of large bicolorable graph states}

\author{Kovid Goyal}
\email[]{kovid@theory.caltech.edu}
\affiliation{Institute for Quantum Information, California Institute of Technology, Pasadena, California 91125, USA}

\author{Alex McCauley}
\email[]{mccaule@caltech.edu}
\affiliation{Institute for Quantum Information, California Institute of Technology, Pasadena, California 91125, USA}

\author{Robert Raussendorf}
\email[]{rraussendorf@perimeterinstitute.ca}
\affiliation{Perimeter Institute, 31 Caroline Street North, Waterloo, Canada, N2L 2Y5}


\date{May 26, 2006}

\begin{abstract}
We describe novel purification protocols for bicolorable graph states. The protocols scale efficiently for large graph states. We introduce a method of analysis that allows us to derive simple recursion relations characterizing their behavior as well as analytical expressions for their thresholds and fixed-point behavior. We introduce two purification protocols with
high threshold. They can, for graph degree 4, tolerate  1\% (3\%) gate error or 20\% (30\%) local error.
\end{abstract}

\maketitle

\section{Introduction}

The known protocols in quantum information processing require a certain degree of
quantum-mechanical entanglement to achieve an
advantage over their classical counterparts. Often, this
quantum-mechanical ``essence'' is provided in terms of in-advance-prepared quantum states.
For example, Bell states are used in a
well-known protocol for quantum cryptography \cite{Ekert1991}, and schemes
for multiparty cryptographic tasks using Greenberger-Horne-Zeilinger
(GHZ) states and other Calderbank-Shor-Steane (CSS) states have been
devised \cite{Chen2004}. Further, in quantum computation, multiparticle
entangled states can be used to streamline the execution of gates and
subcircuits via gate teleportation \cite{Gottesman1999}, and cluster states
represent a universal resource for quantum computation by local
measurements \cite{Raussendorf2001}.

In most realistic scenarios the quality of entangled resource states
is degraded by the effects of decoherence and methods of error
detection or correction are required to counteract this process. One such method is
state purification where a (close to) perfect copy of a
quantum state is distilled out of many imperfect ones. Purification
was first described for Bell states \cite{Bennett1996a,Bennett1996,Lo1999} and
subsequently generalized to bicolorable graph states and
CSS states  \cite{Dur2003, Aschauer2005,Hostens2005}. Recently, a protocol for the
purification of $W$ states was presented in \cite{Miyake2005}. State purification
is used, for example, to establish a perfect quantum channel between
two parties \cite{Bennett1996a}, to efficiently create long-range entanglement
via quantum repeaters \cite{Dur1999} or to render certain schemes for
topological fault-tolerant quantum computation universal \cite{Bravyi2005}.

Imperfect initial states are not the only sources of error
for realistic state purification. With the exception of certain schemes of
topological quantum computation such as \cite{Bravyi2005}, errors in the
gates for purification also need to be taken into
account.

What can we expect to gain from an imperfect purification
procedure? In the process of purification the errors of the initial
state are replaced by the errors of the purifying gates.  Thus, the
amount of error may be reduced if the quality of the initial states is
low compared to the quality of the gates for purification (but above
threshold). Further, purification can be used to {\em{condition}} the
error of a quantum state. For example, imperfect Bell-state
purification can be used to establish a perfectly private if imperfect
quantum channel \cite{Aschauer2002}. In
a multiparty scenario, for some protocols the purification
gates act locally on each copy of the state to purify, resulting in a
local or close to local error model for the final state. This feature
attains relevance in the
context of fault-tolerant quantum computation. Threshold theorems have
been established for increasingly general types of error including
coherent and long-range errors \cite{Terhal2005,Aliferis2006} but there are realistic
scenarios in which standard error correction appears to fail
\cite{Klesse2005}. In such a situation, state purification may be used to turn
the error model into a more benign one.

The focus of this paper is purification of bicolorable graph states by imperfect means,
a subject that has previously been studied in \cite{Aschauer2005,Dur2006,Kruszynska2005}. We are interested in the interplay
between threshold and overhead. Specifically, we seek protocols that,
(I) work with erroneous purification gates, (II) have a high threshold
and good quality of the output state, (III) scale efficiently, and
(IV) are analytically tractable.

Hashing \cite{Chen2004,Bennett1996,Hostens2005} protocols have a high threshold in the error of the initial
state and require only a minimal resource overhead,
but they break down as soon as the purification gates become
slightly imperfect \footnote{For hashing, all $N$ copies are included from the beginning. Each qubit of the state copies which are later measured is acted upon by a
large number of noisy CNOT-gates.  The error-correction procedure is applied only after the CNOTs have acted, such that their errors accumulate. Thus in the large $N$ limit no matter how small the gate noise, the output state will be severely affected and the protocol will fail.}. Recursive protocols such as \cite{Dur2003} also have a
high threshold for error in the initial states and furthermore work
with imperfect purification gates, but they are exponentially inefficient
in the number of particles.

Our protocols are resistant to initial as well as purification errors and are
computationally efficient. As a
bonus, our protocols are analytically tractable for a wide
class of errors. Specifically, our base protocol described in
\Sref{sec:3copy} can be analyzed for arbitrary input states and general probabilistic
Pauli errors in the purification gates. This fact arises through a special locality
property. So far, the exponential
increase of parameters in the
description of  $n$-particle mixed states---even mixed stabilizer
states---has been found to be an obstacle to analytic discussion, and only
severely restricted error models have been treated in the literature.

This paper is organized as follows: in \Sref{sec:post-selection} we briefly review
the protocol \cite{Dur2003} for purification of bicolorable graph
states. In Secs. \ref{sec:3copy} and \ref{sec:conditional-bandaid} we describe our purification protocols
and characterize them in terms of purification threshold, output
quality, and overhead. We conclude with a discussion of our results in
Section \ref{sec:conclusion}.

\section{Brief review}\label{sec:post-selection}
Consider a graph $G(V, E)$ with vertex set $V$ and edge set $E$. $G(V,E)$ is bicolorable if $V$ can be partitioned into two disjoint subsets $A$ and $B$ such that every edge in $E$ connects a vertex in $A$ with a vertex in $B$. $E$ defines a \textit{neighborhood} relation on elements of $V$; $\neigh(j):=\lbrace i \in V : (i\;j) \in E\rbrace$. Define the correlation operators
\begin{equation}
\gen_j := \X_j \prod_{i\in\neigh(j)} \Z_i
\end{equation}
 where $\X$, $\Y$, and $\Z$ are the Pauli matrices. A graph state is a \vectornorm{V}-qubit state $\ket{\Mu}$ ($\Mu \in \left\lbrace0, 1\right\rbrace^{\vectornorm{V}}$) that satisfies the eigenvalue equations
\begin{equation}\label{eq:eigenval}
\gen_j \ket{\Mu} = (-1)^{\mu_j}  \ket{\Mu},  \forall j=1,\dots,\vectornorm{V}.
\end{equation}
The states $\left\lbrace\ket{\Mu}\right\rbrace$ form a basis of the Hilbert space of \vectornorm{V}-qubit states called the \textit{graph basis}.

\begin{figure}
\includegraphics{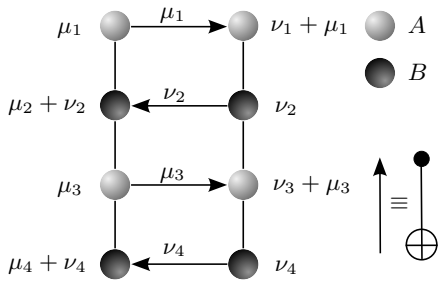}
\caption{\label{fig:mcnot}Action of MCNOT in the graph basis. The arrows represent the direction of syndrome (or $\Z$ error) flow (i.e. the action of the MCNOT on the stabilizer)}
\end{figure}

We now briefly discuss the post-selection protocol of \cite{Dur2003}. The protocol works by taking two identical copies of a bicolorable graph state and performing multiple CNOTs (MCNOT) between them, in a definite pattern as illustrated in \Fref{fig:mcnot}.
Relabeling states in the graph basis to reflect the partition into colors $A$ and $B$ (i.e., $\ket{\Mu}\equiv\ket{\Mu_A,\Mu_B}$), the effect of the MCNOT is \cite{Dur2003}
\begin{equation}\label{eq:mcnot}
\ket{\Mu_A,\Mu_B}\ket{\Nu_A,\Nu_B}\mapsto\ket{\Mu_A,\Mu_B+\Nu_B}\ket{\Nu_A+\Mu_A,\Nu_B},
\end{equation}
where $+$ is elementwise addition modulo 2. Notice that information about $\Mu_A$ has been copied into state 2 and information about $\Nu_B$ has been copied into state 1. We then measure the local observables $\X$ and $\Z$ on copy 2, and reconstruct from the measurement outcomes the eigenvalues of all $\gen_j$ with $j \in \A$. Suppose we get -1 at the $k$th qubit. Then we know that either $\mu_k$ or $\nu_k$ was 1, but we do not have enough information to decide which one, so we throw away the states and start again. We keep doing this until all measurements are clear. By this procedure we correct, to lowest order, errors in the qubits of color $A$. In the next round we interchange the roles of colors $A$ and $B$ and so purify the $B$ qubits. We can concatenate this procedure to achieve desired levels of purity. Because we are post-selecting states on the basis of a global measurement outcome, this protocol is inefficient for large states. This inefficiency can be addressed by using error correction instead of post-selection, to which we now turn.

\section{Three-copy protocol}\label{sec:3copy}
\begin{figure*}
\includegraphics{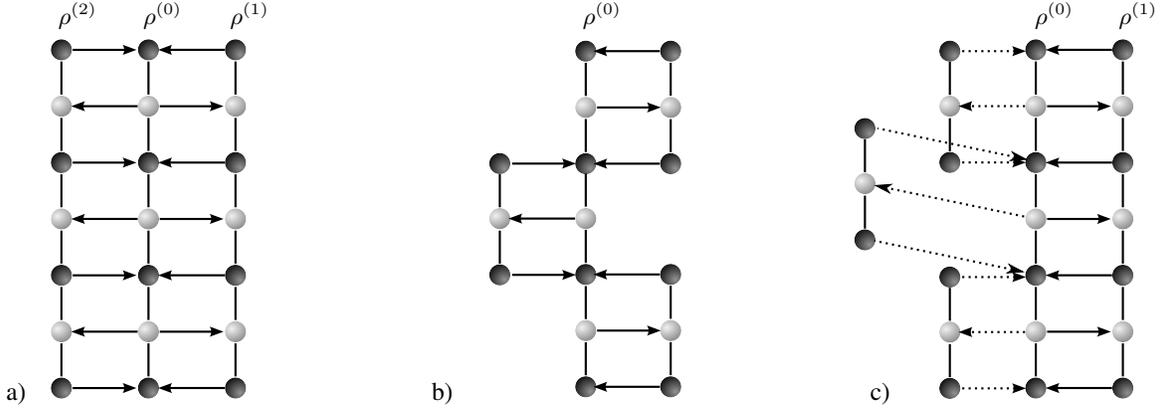}
\caption{\label{fig:protocols}The MCNOT for subprotocol $P_1$ in (a) The 3-copy protocol, (b) The band-aid protocol and (c) The conditional bandaid protocol. The dotted lines in c) indicate that the bandaids are applied only if there is an ambiguous syndrome at that location. Here we show graphs of degree 2, but these protocols can be applied to graphs of any degree.}
\end{figure*}

The simplest way to get enough information to perform error correction is to do the MCNOT on three copies instead of two. The three-copy protocol consists of two subprotocols. We use three identical copies of the state in each subprotocol. The output of the first subprotocol is used as input for the next. Thus, we need nine copies to run a single round. Let the three identical copies be \zero, \one, and \two. Subprotocol 1 ($P1$):
\renewcommand{\labelenumi}{\roman{enumi}.}
\begin{enumerate}
\item Partition the graph into two colors \A and \B ($V=V_\A\cup V_\B \text{ and } V_\A\cap V_\B=\varnothing$).
\item Perform the MCNOT between copies \zero and \one and \zero and \two such that information about qubits of color \A flows from $\zero \rightarrow \one$ and $\zero \rightarrow \two$. As a side effect information about \B will flow from $\one,\two \rightarrow \zero$. See \Fref{fig:protocols}(a), below.
\item \label{steps:3copy:3}Measure qubits of color \A in the $\X$ basis and qubits of color \B in the $\Z$ basis in states \one and \two. This is a measurement of $\gen_j$ for $j \in \A$. If the measurement of $\gen_j$ gives $+1$ $(-1)$ we get a syndrome of $0$ $(1)$. Thus, for each $j \in \A$ we have two bits of syndrome $\sigma_j^{(1)}$ and $\sigma_j^{(2)}$.
\item Apply the correction $\prod_{j\in \A} \Z_j^{\sigma_j^{(1)}\cdot \sigma_j^{(2)}}$ to \zero.
\end{enumerate}
For subprotocol $P2$ the roles colors $A$ and $B$ are interchanged.

First, we will analyze this protocol with ideal CNOT gates. This will allow us to derive simple closed-form recursion relations characterizing the behavior of the protocol, as well as analytical estimates of the threshold and efficiency. In \Sref{sec:3copy:noisy} we generalize to noisy gates. The analysis is restricted to density matrices that are diagonal in the graph basis (i.e., probabilistic mixtures of graph states). At the end of \Sref{sec:3copy:noisy}, we will show that our results are valid for arbitrary density matrices.
\subsection{Ideal gates}\label{sec:3copy:ideal}
Equation \ref{eq:mcnot} implies that the effect of the MCNOT on \zero, \one, and \two is
\begin{align}\label{eq:3copy:mcnotb}
\ket{\Mu_A^{(0)},\Mu_B^{(0)}}  &\mapsto\ket{\Mu_A^{(0)},\Mu_B^{(0)}+\Mu_B^{(1)}+\Mu_B^{(2)}}\\\notag
\ket{\Mu_A^{(1)},\Mu_B^{(1)}} &\mapsto \ket{\Mu_A^{(1)}+\Mu_A^{(0)},\Mu_B^{(1)}}\\\label{eq:3copy:mcnota}
\ket{\Mu_A^{(2)},\Mu_B^{(2)}} &\mapsto \ket{\Mu_A^{(2)}+\Mu_A^{(0)},\Mu_B^{(2)}}.
\end{align}
Equation \ref{eq:eigenval} implies that the effect of the correction is
\begin{equation}\label{eq:3copy:err_corr}
\ket{\Mu^{(0)}_\A,\Mu^{(0)}_\B}\mapsto\ket{\Mu_\A^{(0)}+\bm{\sigma},\Mu^{(0)}_\B},
\end{equation}
where $\sigma_j:=\sigma_j^{(1)}\cdot \sigma_j^{(2)}$. By measuring \one and \two, we get two bits of syndrome for each qubit of color $A$ in \zero. The syndrome is conclusive; it allows us to identify, to lowest order in the error probability on which state the error occurred. We can thus do error correction instead of post-selection. This will make the protocol scale
efficiently in the size of the states. The price is a reduction of the
threshold value.

We now derive a recursion
relation for the expectation values $\avgj$, $j \in 1,\dots,N$. They yield a
necessary and sufficient condition for purification. For the moment we
assume that the initial state $\rho$ is diagonal in the graph basis--i.e.,
$\rho$ is a probabilistic mixture. It is then safe to consider error
probabilities. This assumption is not necessary, however. It is
removed in \Sref{sec:3copy:noisy}. Define $P_j(\rho)$ as the probability to find the eigenvalue $-1$ in the measurement of $\gen_j$ on $\rho$ as
\begin{equation}\label{eq:pj}
P_j(\rho) := \Tr\left[\frac{1-\gen_j}{2}\rho\right] = \frac{1-\avgj}{2}.
\end{equation}
Consider subprotocol $P1$. In order to analyze this protocol we make use of the fact that the error correction operation is local. It only uses information about \avgj in each copy to apply a correction to the $j$th qubit in \zero. Thus, $\avgj$ should have nice decoupled recursion relations. We will later derive the recursion relations for the expectation value of arbitrary stabilizer elements, which in general are more complex.

First consider qubits of color \B. From \Eref{eq:3copy:mcnotb}, $\mu_j^{(0) } \mapsto \mu_j^{(0)}+\mu_j^{(1)}+\mu_j^{(2)}$. Since our copies are identical, we have $P_j(\zero)=P_j(\one)=P_j(\two)=P_j$. Then, $P_j \mapsto P_j^3 + 3P_j(1-P_j)^2$. In terms of expectation values,
\begin{equation}\label{eq:3copy:b}
\avgj'= \avgj^3.
\end{equation}
Under concatenation of $P1$ with itself, qubits of color \B are \textit{polluted} with $\avgj_\zero \rightarrow \avgj_{\I}=0$.

\begin{figure}[htp]
\includegraphics{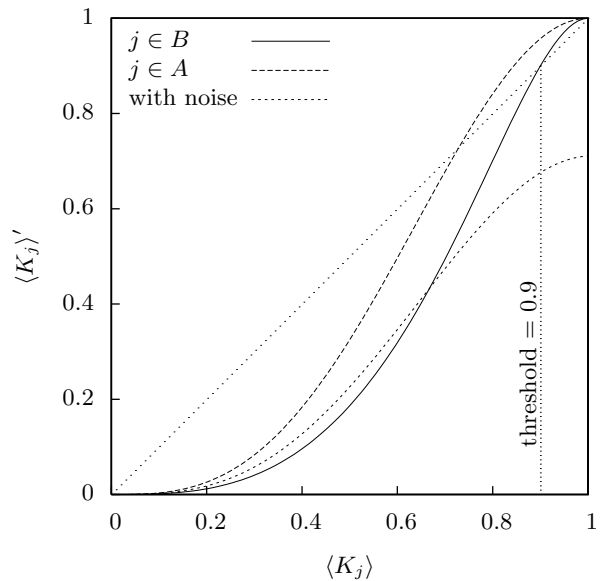}
\caption{\label{fig:3copy:rec}Recurrence curves for the three-copy protocol. These simple curves fully encapsulate the behavior of the protocol with ideal gates. The point of intersection with $\avgj'=\avgj$ gives the threshold. If the gates are too noisy, the protocol breaks down, as indicated by the lowest curve.}
\end{figure}

Turning our attention to qubits of color \A we note that error correction fails if $\mu_j =1$ for more than one copy. Thus,  $P_j \mapsto P_j^3 + 3P_j^2(1-P_j)$. In terms of expectation values
\begin{equation}\label{eq:3copy:a}
\avgj' =  \frac{1}{2}\left(3-\avgj^2\right)\avgj.
\end{equation}
Under concatenation of $P1$ with itself, qubits of color \A are \textit{purified} with $\avgj_\zero \rightarrow  \avgj_{\ket{\vec{0}}\bra{\vec{0}}}=1$.

Subprotocol $P2$ is identical to $P1$ except that the roles of \A and \B are interchanged and the three copies are the output states from running $P1$ 3 times. The three-copy protocol is the composition of $P2$ with $P1$. Let $P=P2\circ P1$; then Eqs. (\ref{eq:3copy:b}) and (\ref{eq:3copy:a}) imply that under the action of $P$
\begin{equation}\label{eq:3copy:rec}
\avgj' = \begin{cases}
                  \frac{1}{8}\left(3-\avgj^2\right)^3\avgj^3& \text{if $j\in\A$},\\
                  \frac{1}{2}\left(3-\avgj^6\right)\avgj^3& \text{if $j\in\B$}.
          \end{cases}
\end{equation}

The recursion relations (\ref{eq:3copy:rec}) have, for each color, a unique repulsive
fixed point in the interval $(0,1)$ which separates the basins of attraction
for the trivial fixed point at 0 and the nontrivial fixed point at 1 (See \Fref{fig:3copy:rec}). The upper
fixed point corresponds to the perfect graph state. Thus, the
stated protocol purifies a graph state if and only if
\begin{align}\label{eq:3copy:thresholds}\notag
  \avgj &> 0.7297\text{ for all $j$ in $\A$}\\
  \avgj &> 0.9003\text{ for all $j$ in $\B$}.
\end{align}
We can compare these thresholds to the thresholds for the post-selection protocol of \cite{Dur2003}. For this protocol, it is not known how to derive a threshold for general noise or even probabilistic Pauli noise. However, for the particular case where only independent local phase flip errors are assumed for the
initial states, recursion relations can be derived even for post-selection. Then, the P1 (post-selection) recursion relation for \avgj with $j \in \B$ is $\avgj'=\avgj^2$ and for $j\in\A$ is
$\avgj'=\frac{2\avgj}{1+\avgj^2}$. The resulting threshold values are $\avgj_{\text{th}}=0.2956$ for $j\in\A$ and $\avgj_{\text{th}}=0.5437$ for $j \in \B$.

Returning to our protocol, it is possible to derive recursion relations for the expectation values of arbitrary stabilizer elements. They are not in general decoupled, but there is still a notion of locality.
The generalized relation allows us to compute the recursion relations for stabilizers with small support efficiently. Define
\begin{equation}\label{eq:gen_stab_defn}
\gen_{\vec{a},\vec{b}} := \prod_{i=1}^{\vectornorm{V_\A}}\gen_i^{a_i}\prod_{j=1}^{\vectornorm{V_\B}}\gen_j^{b_j},
\end{equation}
where $\vec{a}\in\bin^{\vectornorm{V_\A}} \text{ and } \vec{b}\in\bin^{\vectornorm{V_\B}}$. The factors in the first product are the stabilizer generators for qubits of color \A, while those in the second product are for qubits of color \B. Then (see \Aref{sec:generalized_rec}) under the action of subprotocol $P1$,
\begin{widetext}
\begin{equation}\label{eq:gen_stab_rec}
\avg{\vec{a},\vec{b}}' = \frac{1}{2^{\vectornorm{\vec{a}}}} \sum_{\vec{a}_1,\vec{a}_2\ll\vec{a}} (-1)^{\vec{a}_1\cdot\vec{a}_2} \avg{\vec{a}+\vec{a}_1+\vec{a}_2,\vec{b}}\avg{\vec{a}_1,\vec{b}}\avg{\vec{a}_2,\vec{b}},
\end{equation}
\end{widetext}
where $\vec{f}\ll\vec{g}\text{ iff } f_j = 0 \text{ whenever }g_j =0$. Equations~(\ref{eq:3copy:a}) and (\ref{eq:3copy:b}) are special cases for $\avg{\vec{a},\vec{b}}=\avgj$ with $j\in\A,\B$ respectively. An interesting feature of this equation is that it relates a correlator of weight $w = \vectornorm{\vec{a}}+\vectornorm{\vec{b}}$ to correlators of weight no more than $w$. This makes it feasible to calculate the recursion relations for correlators of small weight.

In order to discuss the behavior of this protocol under concatenation with itself, it is useful to switch back to probability variables. Then \Eref{eq:3copy:rec} implies that if the protocol is concatenated with itself $k$ times,
\begin{equation}\label{eq:3copy efficiency}
P_j(\rho(k))\leq \left(\frac{P_j(\rho(0))}{P_{\text{th}}}\right)^{2^k}
\end{equation}
where $P_{\text{th}}$ is the threshold error probability. The $k$-concatenated protocol requires $3^{2k}$ identical copies, thus, the protocol is exponentially efficient under concatenation. The reduction of error, \Eref{eq:3copy efficiency}, is not conditioned on a particular post-selected
syndrome. The overhead in number of required initial states is independent of the size $N$ of the graph state. We conclude that under concatenation the protocol reaches the reference state $\ket{\vec{0}}$ with efficient use of resources. Contrarily, for the post-selection protocol \cite{Dur2003} the overhead acquires a dependence  $\exp(\alpha N)$, with some $\alpha>0$, due to post-selection of a particular syndrome.

\subsection{Noisy gates}\label{sec:3copy:noisy}
Now we investigate what happens to this protocol when the CNOT gates themselves are noisy. In the three-copy protocol CNOT gates act on the same qubit in two states $\rho^{(m)}$ and $\rho^{(n)}$. We model a noisy two-qubit gate as an ideal gate followed by the two-qubit depolarizing channel [i.e., the $SU(4)$-invariant channel]
\begin{equation}\label{eq:depol}
\T^{(k)} := (1-p_2)[\I]+\frac{p_2}{16}\sum_{i,j=1}^4\left[D_i^{(k,m)}\otimes D_j^{(k,n)}\right],
\end{equation}
where $D_{i,j}\in\lbrace\I,\X,\Y,\Z\rbrace$ and $k$ is the qubit index. $D^{(k,m)}$ acts on the $k$th qubit of $\rho^{(m)}$. The $\Z$ gates applied in the error-correction steps and the measurement of the syndrome are assumed to be noiseless. This is natural since the Pauli phase flips $\Z$ may be omitted as physical operations and instead accounted for in the classical syndrome processing. We will include the effect of measurement errors in the analysis when we consider the more sophisticated protocols, which have higher thresholds than the three-copy protocol. If we consider the effect of $\T^{(k)}$ only on \avgj in state \zero, then using \Eref{eq:eigenval} we can reduce the noise to an effective error. For every $k\in V:k\in\neigh(j)\cup\lbrace j\rbrace$
\begin{equation}\label{eq:flipNoise}
\T_{\text{eff}}^{(k,j)}(\zero)=\left(1-\frac{p_2}{2}\right)[I]+\frac{p_2}{2} [Z_j^{(0)}].
\end{equation}
If $k\notin\neigh(j)\cup\lbrace j\rbrace$, then $\T_{\text{eff}}^{(k,j)}$ is just the identity map. Since every error channel commutes with every CNOT, we can model the noisy MCNOT as the ideal MCNOT followed by $\vectornorm{V}$ noise channels.

The error channel \Eref{eq:depol} is local (i.e., it acts only on qubit $k$ in $\rho^{(m)}$ and $\rho^{(n)}$). Also the error operators are Pauli operators, which map graph states to graph states, keeping $\rho$ diagonal in the graph basis. Thus we can expect the noisy recursion relations to have the same form as \Eref{eq:3copy:rec}. Considering only subprotocol $P1$, the $j$th qubit in \zero is affected by $2(d+1)$ error channels. For simplicity, we assume all vertices of the graph have the same degree $d$. If this is not the case, then there would be a different set of recursion relations for each degree. We can then choose $d$ to be the maximum degree, in which case the recursion relations will be lower bounds for all other degrees. The total probability that the $j$th qubit is flipped by an error is $\frac{1-(1-p_2)^{2(d+1)}}{2}$. Thus, for qubits of color $B$,
\begin{equation}\label{eq:3copyn:b}
\avgj'=\alpha^2 \avgj^3,
\end{equation}
where $\alpha=(1-p_2)^{(d+1)}$.

The situation is a little more complex for qubits of color $A$ as the error in the MCNOT between \zero and \one is propagated by the MCNOT between \zero and \two (see \Fref{fig:protocols}(a)). However, the form of the recursion relation remains the same. We get
\begin{equation}\label{eq:3copyn:a}
      \avgj'=\frac{\alpha^2}{2}\left(2+\alpha^{-1}-\avgj^2\right)\avgj.
\end{equation}
For a derivation see \Aref{sec:recurrence:noisy}. Composing subprotocols $P1$ and $P2$ we get the recursion relations for the three-copy protocol with noisy gates
\begin{equation}\label{eq:3copy:noisy}
\avgj' = \begin{cases}
                  \frac{\alpha^8}{8}\left(2+\alpha^{-1}-\avgj^2\right)^3\avgj^3& \text{if $j\in\A$},\\
                  \frac{\alpha^4}{2}\left(2+\alpha^{-1}-\alpha^4\avgj^6\right)\avgj^3& \text{if $j\in\B$},
          \end{cases}
\end{equation}
Here, qubits of color \A behave worse. Solving the recursion relations for fixed points, we find that there are two non-trivial positive fixed points (see \Aref{sec:fixed:points}) for $\alpha > 0.9902$. Consider the interval $[0,1]$. It has at most three fixed points $0=f_0<f_1\leq f_2\leq 1$. $f_0$ and $f_2$ are attractive while $f_1$ is repulsive. Thus $f_2$ will be a stable fixed point for $\alpha > 0.9902$ and $\avgj_{\text{initial}}>f_1$. This gives a threshold for the noise affecting the gates that scales inversely proportional to the graph degree $d$,
\begin{equation}
  p_{\text{th}}\approx\frac{9.8\times10^{-3}}{d+1}.
\end{equation}
Specifically for degrees 2 and 4 we obtain
\begin{equation}
p_\text{th}=\begin{cases}
0.328\;\%\text{ for $d=2$},\\
0.197\;\%\text{ for $d=4$}.
\end{cases}
\end{equation}
This is a rather low value, but it will be substantially improved when we consider more sophisticated protocols.

We now show that the recursion relations \Eref{eq:3copy:noisy} are
valid regardless of whether or not the considered states are diagonal in the
graph basis. To see this, let us define a depolarization operator
\D which converts an arbitrary $n$-qubit mixed state $\rho$ into an
$n$-qubit mixed state $\rho_D = \D \rho$ that is diagonal in
the graph basis. $\D$ takes the form
\begin{equation}
  \label{eq:depolOp}
  \D = \left(\prod_{\ab} \frac{[I]+[\gen_{\ab,\vec{0}}]}{2}\right) \left(\prod_{\bb} \frac{[I]+[\gen_{\vec{0},\bb}]}{2}\right),
\end{equation}
where $\ab$ and $\bb$ are vectors in a basis of $\bin^{\vectornorm{V_{\A}}}$ and $\bin^{\vectornorm{V_{\A}}}$ respectively.

We only consider $P1$, the first round of the protocol. It is
associated with a transformation $P1:\, \rho \longrightarrow \rho^\prime=\R\left(\rho^{\otimes 3}\right)$.
$\R$ and \D commute--i.e.,
\begin{equation}\label{eq:comm}
  \R\left((\D\rho)^{\otimes 3}\right) = \D \circ \R\left(\rho^{\otimes 3}\right),
\end{equation}
for any $\rho$. For a proof see \Aref{app:general}.

Consider a recursion relation of the form
\begin{equation}\label{eq:rec01}
  \expec{\gen_{\vec{a}, \vec{b}}(\rho_D^\prime)} =  f_{\vec{a},\vec{b}}\left(\{\expec{\gen_{\vec{i},\vec{j}}(\rho_D)}\}\right),
\end{equation}
with $f_{\vec{a},\vec{b}}$ some function depending on
$\vec{a},\vec{b}$ as in \Eref{eq:gen_stab_rec}. Now,
\begin{align*}
  \expec{\gen_{\vec{a},\vec{b}}(\rho_D^\prime)} &= \Tr\left[\gen_{\vec{a},\vec{b}}\R\left(\left(\D\rho\right)^{\otimes3}\right)\right]&\\
  &= \Tr\left[\gen_{{\vec{a}},{\vec{b}}} \D \circ \R\left(\rho^{\otimes3}\right)\right]&\text{[by \Eref{eq:comm}]}\\
  &=\Tr\left[\D^\dagger(\gen_{{\vec{a}},{\vec{b}}})\rho^\prime \right]&\text{(trace cyclicity)}\\
  &=\langle \gen_{{\vec{a}},  {\vec{b}}}(\rho^\prime)\rangle.&\text{($\D^\dagger\equiv\D$)}
\end{align*}
Similarly, $\expec{\gen_{{\vec{i}},{\vec{j}}}(\rho_D)}
= \expec{\gen_{{\vec{i}},{\vec{j}}}(\rho)}$, such that
\begin{equation}
   \expec{\gen_{{\vec{a}}, {\vec{b}}}(\rho^\prime)} = f_{{\vec{a}},{\vec{b}}}\left(\{\expec{\gen_{{\vec{i}},{\vec{j}}}(\rho)}\}\right).
\end{equation}
Thus, a recursion relation of the form of \Eref{eq:rec01} such as \Eref{eq:3copy:noisy} holds
for all states $\rho$ and not just for diagonal states $\rho_D=\D\rho$.

\section{Improved Protocols}
\subsection{Error model}\label{sec:error model}
In the following, we consider a scenario where graph states are created locally from product states, then distributed to several parties and subsequently purified. Errors occur in each of these steps--specifically, the following
\begin{itemize}
  \item There is a two-qubit error $T$, \Eref{eq:depol}, associated with each controlled-PHASE gate in the creation of the graph state, with probability $p_2$.
  \item A local depolarizing error with probability $p_1$ occurs on each graph state qubit during transmission.
  \item Every CNOT gate used in purification carries a two-qubit error, \Eref{eq:depol}, with error probability $p_2$. Every measurement is modeled by a one-qubit depolarizing channel with error probability $p_2$ followed by a perfect measurement.
\end{itemize}

\subsection{Bandaid protocol}
In order to raise the threshold of the three-copy protocol, we will try to combine the strategies of error correction and post-selection (which has a higher threshold). One way to do this is to use small highly purified GHZ states--i.e., bandaids, to purify the graph one vertex at a time. The usual MCNOT is performed between the bandaid and the large graph state as shown in \Fref{fig:protocols}(b). This copies information about the central vertex into the bandaid which is then measured to give a syndrome. Since the bandaid is highly purified (for example, by post-selection), it does not pollute the large state much. It is important to note that the error correction is still local, and we expect the recursion relations to be decoupled as in the case of the three-copy protocol.

\renewcommand{\labelenumi}{\roman{enumi}.}
The bandaid protocol also has two subprotocols. The first one $P1$ is the following.
\begin{enumerate}
       \item Partition the graph into two colors \A and \B ($V=V_\A\cup V_\B \text{ and } V_\A\cap V_\B=\varnothing$).
       \item The bandaids are placed over the large state such that each central qubit of the bandaid is over a vertex of qubit \A for all qubits of color \A. Perform the MCNOT as shown in \Fref{fig:protocols}(b).
       \item Measure the central qubit of each bandaid in the $\X$ basis and the other qubits in the $\Z$ basis. For each bandaid multiply the measured eigenvalues. If the product is (-1) +1 then the syndrome bit $\sigma_j$ is (1) 0.
       \item Apply the correction $\prod_{j\in\A}Z_j^{\sigma_j}$ to the large state.
\end{enumerate}
$P2$ is the same as $P1$, with the roles of colors $A$ and $B$ reversed.

Consider subprotocol $P1$. For qubits of color \B the argument is very similar to the three-copy protocol, except that each qubit is affected by two gates from each of $d$ bandaids. Thus,
\begin{equation}\label{eq:bandaid:b}
       \avgj'=(1-p_2)^{2d}\avgj\avgj_b^{d},
       \end{equation}
where $\avgj_b$ is the constant initial purity of the bandaid.

For qubits of color \A, first suppose that the CNOT gates are ideal. Then, a simple transfer of purity occurs:
\begin{equation}\label{eq:bandaid:ideal}
	\avgj' = \avgj_b.
\end{equation}
If the gates are noisy, \Eref{eq:bandaid:ideal} is multiplied by a noise factor of the form $(1-p_2)^{f(d)}$ as in the case of the three-copy protocol. There is a subtlety involving the temporal ordering of the bandaids. The bandaids do not all commute with each other. There are $1+d(d-1)$ bandaids that affect qubit $j$. One of them is the bandaid that is used to purify the qubit. On average $k=\frac{d(d-1)}{2}$ of the rest will be applied before the purifying one. Any effect from the $k$ prior bandaids will be erased by the purifying bandaid [see \Eref{eq:bandaid:ideal}]. The purifying bandaid has $d+1$ noisy CNOTs affecting \avgj; since the noisy MCNOT is modeled as an ideal MCNOT followed by noise, no information about the noise is propagated to the bandaid. Thus, the noise will commute with the error correction procedure. Since a measurement error that flips the central qubit of the bandaid will cause us to apply the wrong error correction operator, it can also be reduced to an effective error as given by \Eref{eq:flipNoise}. Thus, $f(d) = 2(d+1) + k$ and we have
\begin{equation}\label{eq:bandaid:a}
       \avgj' =(1-p_2)^{\frac{d(d+3) +4}{2}}\avgj_b.
\end{equation}
Combining subprotocols $P1$ and $P2$, we get the recursion relations for the bandaid protocol with noisy gates as well as noisy measurements
\begin{equation}\label{eq:bandaid final}
       \avgj'=\begin{cases}
                        (1-p_2)^{\frac{d(d+7)+4}{2}}\avgj_b^{d+1}&\text{ for }j\in\A,\\
                        (1-p_2)^{\frac{d(d+3)+4}{2}}\avgj_b&\text{ for }j\in\B.
                        \end{cases}
\end{equation}
The behavior of qubits of color \A is worse, and we will use their purity as the final purity of the large state.

\begin{figure}[htp]
\includegraphics{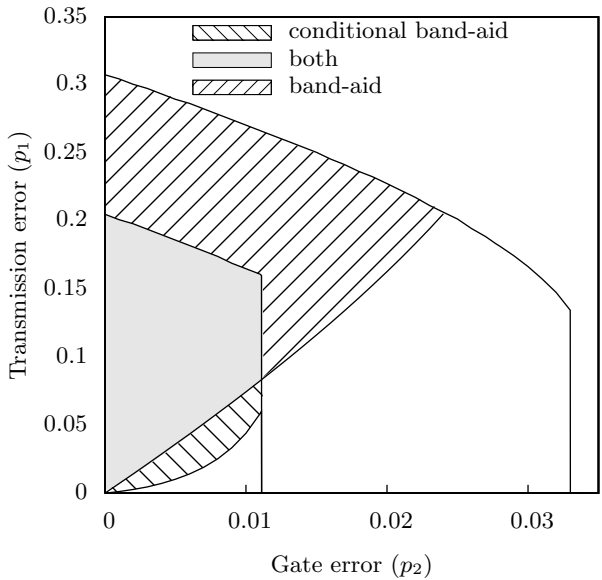}
\caption{\label{fig:bandaid-tradeoff}Trade-off curves for the bandaid and conditional bandaid protocols ($d=4$). The decreasing curves represent the breakdown of the post-selection protocol, when there is too much error. The increasing curves demarcate the region where the final purity of the purified states is higher than the purity of the unpurified states. It makes sense to purify in the shaded regions.}
\end{figure}

As per our error model in \Sref{sec:error model}, the noisy CPHASE, CNOT and measurement gates are parametrized by $p_2$. The noisy transmission channel is parametrized by $p_1$. For the final result, we need to know the quality of the bandaids. We assume that these are also created locally, then transmitted and purified. The bandaids, however, are of fixed size and may thus be purified by the post-selection protocol \cite{Aschauer2005} with the higher threshold. The output quality of the purified bandaids is, to leading order in $p_2$,
\begin{equation}\label{eq:linear postselection}
  \avgj_b = 1- (d+1)p_2,
\end{equation}
such that
\begin{equation}\label{eq:linear bandaid}
  \avgj = 1- \frac{1-d(3d+11)+6}{2}p_2,
\end{equation}
for small $p_2$ [from Eqs.~(\ref{eq:bandaid final}) and (\ref{eq:linear postselection})]. As \Eref{eq:linear bandaid} shows, with increasing graph degree the effect of errors in the purification process is strongly enhanced. One may therefore ask the question whether it is useful to purify at all or whether the transmitted state should be used right away. To decide this we compute \avgj after graph-state creation and transmission,
\begin{equation}
  \avgj = (1-p_2)^{\frac{d(d+1)}{2}}(1-p_1)^{d+1}.
\end{equation}
See \Aref{sec:creation} for a derivation. We compare this expression with \Eref{eq:bandaid final} and find that there is indeed a parameter region where it makes sense to purify. This region is displayed for graphs of degree $d=4$ in \Fref{fig:bandaid-tradeoff}. It is bounded from above and right by the curve which indicates the breakdown of the bandaid purification according to the post-selection protocol \cite{Aschauer2005}. If we use post-selection to obtain bandaids of high purity, then the threshold of the bandaid protocol for degree $d$ graph states equals the threshold for purification of a $d+1$-qubit GHZ state with the post-selection protocol [8]. However, the output purity of the bandaid protocol is smaller. Only above the ascending curve is it advantageous to purify.
\subsection{Conditional bandaid protocol}\label{sec:conditional-bandaid}
In order to correct the $d^2$ dependence of the fixed point in the bandaid protocol, we will combine it with the three-copy protocol. The hybrid protocol, called the conditional bandaid protocol, sacrifices in threshold to improve the fixed point. The fixed-point behavior, at least to linear order in gate noise, is almost as good as that of the post-selection protocol.

This protocol proceeds in the same fashion as the three-copy protocol, except that two copies are used per round, and wherever a measurement of $\gen_j$ yields eigenvalue $-1$ (i.e., an error), a post-selected bandaid is applied to purify qubit $j$ (see \Fref{fig:protocols}). For small gate noise, we expect to have to apply only a few bandaids per round, nonetheless, the threshold is set by the qubits to which we have to apply bandaids. Locations where a measurement of $\gen_j$ yields $1$ are error free to lowest order. Once again, we have two subprotocols, $P_1$ and $P_2$ each purifying a different color.

The analysis is similar to that used in arriving at \Eref{eq:3copy:noisy} for the three-copy protocol. However, the situation is complicated by the fact that the bandaids are applied conditioned on the results of measuring \one. As a result, the recursion relations for the one point correlators are no longer completely decoupled. We can, however, find a simple lower bound on them.

Define $\avg{b}$ to be the minimum purity of the post-selected bandaid. It is a constant. For simplicity we assume that all qubits in the bandaid have this purity. As before, we assume that the graph of the large state is translationally invariant, i.e., all vertices have the same degree.  The definition $\beta \equiv (1-p_2)^2 \avg{b}$ will be useful. Consider qubits of color $A$ in subprotocol $P_1$, then, by a derivation similar to \Eref{eq:3copyn:a},
\begin{equation}
\avg{j}' =  \frac{\alpha}{2}\left( 2 \alpha \avg{j} + \avg{b} - \alpha \avg{b} \avg{j}^2 \right),
\end{equation}
where $\alpha=(1-p_2)^{d+1}$ as before.

So far, we have been exact. Now consider subprotocol $P_2$. Again focus on qubits of color $A$. Break $P_2$ down into two steps. In step one, we apply the MCNOT to \zero and \one.  It can be readily verified that $\avg{j} \mapsto \alpha \avg{j}^2$.  In step two, bandaids conditioned on the measurement outcome are applied to qubits of color $B$. Let $\vec{y} \in \{0,1\}^{d}$ be the measurement results for the neighbors of qubit $j$. A measurement result of one means a bandaid must be applied at that location. If a bandaid is applied to a neighbor of $j$, \avgj is affected by the errors on the bandaid, characterized by $\avg{b}$ and by two noisy CNOTs. Thus $\avgj \mapsto \beta^{\vectornorm{\vec{y}}}\avgj$. Summing over measurement outcomes and including step one, we get
\begin{equation}
\avg{j}' =  \alpha \left( \sum _{k=0}^{d} \left( \sum _{\vectornorm{\vec{y}}=k} q_{\vec{y}} \beta ^k\right) \right) \avgj^2,
\end{equation}
where $q_{\vec{y}}$ is the probability of measurement outcome $\vec{y}$. Unfortunately, $q_{\vec{y}}$ is a function of the general stabilizer expectation values $\avg{\vec{a},\vec{b}}$, so we will resort to finding a lower bound. Since $q_0 = 1 - \sum_{\vec{y} \neq 0} q_{\vec{y}}$, we can rewrite the above equation as
\begin{align*}
\avgj' &= \alpha_a \left( (1 - \sum_{\vec{y}\neq 0}q_{\vec{y}}) + \sum_{k >0}^{d} \sum_{\vectornorm{\vec{y}}=k} q_{\vec{y}}\beta^k \right) \avgj^2\\
&\geq \alpha \left( 1 - (1-\beta^d) \sum_{\vec{y} \neq 0} q_{\vec{y}} \right)
\avgj^2,
\end{align*}
using $\beta \leq 1$ to arrive at the inequality.

Now, $q_0$ is just the probability that no error is detected on any of the neighbors
of $j$.  Let $p_j$ be the probability of detecting an error on site $j$.  Then, by definition, $\avg{i} = 1-2 \sum _{y | y_i = 1} q_y$. This implies that $\sum_{\vec{y} \neq 0} q_{\vec{y}} \leq \sum_{i \in \neigh(j)} \frac{1-\avg{i}}{2}$. Putting this into the above inequality,
\begin{equation} \label{eq:cdllowerbnd}
\avgj' \geq \alpha \left( 1 - \frac{d}{2}(1-\beta^d) \left(1 - \avg{i}
\right)\right) \avgj^2,
\end{equation}
where \avg{i} is the purity of qubits of color $B$ from the previous round.

Solving for the fixed point, we get, to leading order in gate noise $p_2$,
\begin{equation}
  \avgj = 1 - 2 (d+1) p_2.
\end{equation}
Comparing this to \Eref{eq:linear postselection}, we see that the fixed-point scaling with degree is almost as good as in the post-selection protocol. We now apply the conditional bandaid protocol to the same situation--of a graph state being shared among widely separated parties, as for the bandaid protocol. The results for a degree four state are plotted in \Fref{fig:bandaid-tradeoff}. We see that the threshold (upper) curve is worse, whereas the fixed-point (lower) curve is better for this protocol, as compared to the bandaid protocol. The total purifiable area is smaller, indicating that it breaks down faster. In some sense, we have traded threshold for fixed point. These conclusions hold for arbitrary degree, and the curves are independent of the size of the state, making this protocol eminently suitable for the purification of large bi-colorable graph states.
\section{Conclusion And Outlook}\label{sec:conclusion}
We have described novel purification protocols for bi-colorable graph states
and discussed their performance. The criteria for our protocols are that they do not break
down in the presence of small amounts of noise in the purification process, that they have
a high purification threshold and good output quality, scale efficiently, and be analytically
tractable.

Our final protocol can, for relevant graph states of degree 4, tolerate $1\%$ gate or $20\%$ local
transmission error. These are about 1/3 and 2/3 of the respective values for the post-selection protocol
\cite{Aschauer2005,Dur2003}. However, in contrast to this reference protocol, our protocol scales efficiently
with the graph size.

All our protocols can be treated analytically. In particular, for the three-copy protocol we
derive closed, exact one-dimensional recursion relations in the appropriate observables, irrespective
of the size of the state.

We would like to comment on the influence of the graph degree for the purification threshold. First note that for the three-copy protocol of Section III, in the case of perfect purification gates, the recursion relations  (\ref{eq:3copy:rec}) are completely independent of the graph structure, and so are the thresholds (\ref{eq:3copy:thresholds}). This behavior changes if noise is included in the purification. The critical noise level per purification gate---at which the protocol breaks down---scales inversely proportional with the graph degree. The unfavorable dependence on the graph degree is present in all three protocols we discuss. Thus, the lesson we learn for the case of noisy purification is to beware of large graph degrees. Large graph degrees occur, for example, in graphs states corresponding to codewords of concatenated CSS codes.

 We would also like to comment on the structure of the nontrivial fixed point in our protocols. In the case of erroneous
purification gates, the nontrivial fixed point is not completely specified by the lowest order
expectation values \avgj and it remains to be discussed which error correlations are removed
by the purification protocol. As a first result in this direction, for the three-copy protocol discussed
in Section III we have shown (in \Aref{sec:3copy:correlations}) that correlations of stabilizer expectation values located on
non-overlapping supports are not introduced by the purification procedure if they are absent initially.
This implies that such correlations are absent in all purified states which end up at the same
fixed point as the perfect state. We show in \Aref{sec:fixed:points} that the fixed point for two-generator correlations with distinct
support is unique, which is enough to establish the result that all states at the fixed point obey the relation
$\expec{\gen_i\gen_j}=\avg{i}\avgj$ for such correlations.

A question of further interest is whether the nontrivial fixed point of the protocol is unique at all levels of
correlations. This would imply $\avg{\vec{i}+\vec{j}}=\avg{\vec{i}}\avg{\vec{j}}$ for all correlations with
distinct supports.

Another question of further interest is whether the described or related protocols may be
used to boost the threshold value for fault-tolerant quantum computation \cite{Nielsen2005,Dawson2006,Aliferis2006a,Varnava2005,Raussendorf2006} based on graph
states.
\begin{acknowledgments}
  We would like to thank John Preskill, Frank Verstraete, Jiannis Pachos, Maarten van den Nest, Eric Hostens, Akimasa Miyake, Wolfgang D{\"u}r, Simon Anders, Hans Briegel, Panos Aliferis and Krittika Kanjilal for useful discussions. K.G. is supported by DOE Grant No. DE-FG03-92-ER40701. R.R. has been supported at Caltech by MURI under Grant No. DAAD19-00-1-0374 and by the National Science Foundation under Contract No. PHY-0456720, and is supported by the Government of Canada through NSERC and by
the Province of Ontario through MEDT. Additional support was provided by the National Science Foundation under Grant No. PHY99-07949 during the workshop ``Topological Phases and Quantum Computation'' at the KITP Santa Barbara, and by the Austrian Academy of Sciences.
\end{acknowledgments}

\appendix
\section{Generalized recursion relations}\label{sec:generalized_rec}
We now derive the generalized recursion relations [\Eref{eq:gen_stab_rec}] for the three-copy protocol. While the method used for this derivation is less intuitive, it yields recursion relations for arbitrary stabilizer elements and can handle noisy gates easily.

\subsection{Noiseless gates}
In order to derive \Eref{eq:gen_stab_rec} we work in the \textit{stabilizer basis}. Because $\zero$ is diagonal and the set $\left\{ \avg{\ab,\bb} \right\}$ where $\vec{a} \in \bin^{\vectornorm{V_A}},\vec{b} \in \bin^{\vectornorm{V_A}}$ forms a complete set of observables, we can write an expansion $\zero = \frac{1}{2^{\vectornorm{V_{\A}}+\vectornorm{V_{\B}}}}\sum _{\ab,\bb} \avg{\ab,\bb} \gen_{\ab,\bb}$.

Consider subprotocol $P_1$, which purifies the \A subgraph. The initial state is $\zero \otimes \one \otimes \two$, which can be rewritten as a sum over $\ab,\bb$ of terms of the form
\begin{multline}\label{eq:stabterm}
\avg{\azero,\bzero} \avg{\aone,\bone} \avg{\atwo,\btwo}\times\\ \kket{\azero,\bzero} \kket{\aone,\bone} \kket{\atwo,\btwo}.
\end{multline}

The protocol is linear, so we track the evolution of each term seperately.  Performing step ({\romannumeral 2}), this term becomes
\begin{multline} \label{eq:cnots}
\avg{\azero,\bzero} \avg{\aone,\bone} \avg{\atwo,\btwo}\times\\
\kket{\azero + \aone + \atwo,\bzero} \kket{\aone,\bzero+\bone}\kket{\atwo,\bzero+\btwo}.
\end{multline}

Now consider step ({\romannumeral 3}). Suppose we get measurement outcomes $\measv{1},\measv{2}$ for the stabilizers in subgraph $A$ on copies $\one,\two$.  Then the resultant state is given by applying the projector
\begin{equation}\label{eq:proj}
\frac{1}{2^{2 \vectornorm{V_A}}} \prod _{j=1}^{\vectornorm{V_A}} [\I] \otimes \left([\I]+(-1)^{\meas{1}_j} \gen _{j}^{(1)} \right) \otimes \left([\I]+(-1)^{\meas{2}_j} \gen _{j}^{(2)} \right).
\end{equation}
All the single-site operators involved commute, so this term is a product of stabilizers in $\bb$ and terms of the form
\begin{equation*}
\left([\I] + (-1)^{\meas{k}_j} \gen _{j}^{(k)} \right) \left( \gen _j^{(k)} \right) ^{a_j^{(k)}}.
\end{equation*}
Here $k = 1,2$.  Discarding $\one,\two$, we perform a partial trace over these systems (recalling that $\kket{\ab,\bb}$ are all traceless except $\gen_{\vec{0,0}} = \I$).  In the above term, only the coefficient of $[\I]$ contributes, which is $(-1)^{\meas{k}_j a_j^{(k)}}$.  Including the stabilier operator, we are left with
\begin{multline} \label{eq:ket}
\kron{\bzero}{\bone} \kron{\bone}{\btwo}\frac{(-1)^{\measv{1} \cdot \aone+\measv{2}\cdot \atwo}}{2^{2 \vectornorm{V_A}}}\times\\\kket{\azero+\aone+\atwo,\bone},
\end{multline}
where $\kron{\vec{p}}{\vec{q}}$ is the Kronecker delta on each component of $\vec{p},\vec{q}$. Note that we must have $\bzero=\bone=\btwo$ or the term is zero.

Now examine the action of the Pauli $[\Z]$ operator in this basis. $[\Z] \stab{a,b} = \Z \stab{a,b} \Z = -1^{k} \stab{a,b}$, where $k=0$ iff $\Z$ and $\stab{a,b}$ commute. Effectively, $\Z$ is a diagonal matrix with entries $\pm 1$. Identical reasoning applies to $\X$ and $\Y$. This will make it very easy to add gate noise into the analysis. It also allows us to say that the net effect of the error-correction step {\romannumeral 4} is to multiply \Eref{eq:ket} by a factor of $(-1)^{ (\measv{1} \times \measv{2})\cdot (\azero + \aone + \atwo)}$, where $ (\vec{p} \times \vec{q})_j \equiv p_j \cdot q_j$.  To simplify the notation, change the basis to $\ab \equiv \vec{\azero+\aone+\atwo}$, $\bb \equiv \bzero$.  Then the term becomes
\begin{equation*}
\kron{\bb}{\bone} \kron{\bb}{\btwo} \frac{(-1)^{\measv{1} \cdot \aone+\measv{2}\cdot \atwo + (\measv{1} \times \measv{2})\cdot \ab}}{2^{2 \vectornorm{V_A}}}\kket{\ab,\bb}.
\end{equation*}
In this notation and ignoring the delta functions, the original coefficient in \Eref{eq:cnots} is $\astab{\ab+\aone+\atwo,\bb} \astab{\aone,\bb}\astab{\atwo,\bb}$.  We will now get conditions under which this term contributes to the coefficient of $\kket{\ab,\bb}$.

Summing over measurement outcomes, the coefficient of $\kket{\ab,\bb}$ is
\begin{multline*}
\astab{\ab+\aone+\atwo,\bb} \astab{\aone,\bb}\astab{\atwo,\bb} \times\\\frac{1}{2^{2\vectornorm{V_A}}}\sum _{\measv{1},\measv{2}}(-1)^{\measv{1}\cdot \aone+\measv{2}\cdot \atwo + (\measv{1}\times \measv{2})\cdot \ab}.
\end{multline*}
The sum can be reexpressed as
\begin{equation*}\label{eq:meassum}
\prod _{j=1}^{\vectornorm{V_A}} \sum_{\lambda^{(1)}_j,\lambda^{(2)}_j=0}^1(-1)^{\meas{1}_j a^{(1)} _j+\meas{2} _j a^{(2)} _j + \meas{1}_j \meas{2}_j a_j}.
\end{equation*}
If $a_j=0$, then the $j$th factor is zero unless $a^{(1)} _j = a^{(2)} _j = 0$, in which case it is 4.  Hence, for the term $\avg{\ab+\aone + \atwo,\bb} \avg{\aone,\bb}\avg{\atwo,\bb}$ to survive the procedure, we must have $\aone,\atwo \ll \ab$.  If this holds, then an overall factor of $4^{\vectornorm{V_A} - \vectornorm{\ab}}$ comes out.
If $a_j=1$, then a straightforward calculation shows that the $j$th factor contributes a factor of $2(-1)^{a^{(1)} _j a^{(2)}_j}$.  The overall numerical factor is thus $\frac{1}{2^{\vectornorm{\ab}}}$.
To get the new value of $\avg{\ab,\bb}$, we simply sum over $\aone,\atwo$ since these and only these will contribute to the support of $\kket{\ab,\bb}$ under P1. This gives \Eref{eq:gen_stab_rec}.

\subsection{Noisy gates}\label{sec:recurrence:noisy}
Adding noise to the gates requires very little additional work. We can rewrite the depolarizing channel on qubit $j$ of copy $k$ as
\begin{align*}
D_j^{(k)} [\rho]&= \frac{1}{2}\left( [\I] + [\Z]_j^{(k)} \right) \frac{1}{2}\left( [\I] + [\X]_j^{(k)} \right)[\rho] \\&\equiv P_Z^{(j,k)} P_X^{(j,k)}.
\end{align*}
It was shown above that $[\X],[\Z]$ have $\pm1$ on the diagonal. Thus writing the noise channel in this form illustrates how the noise components act as projectors  $P^{(k)}_{Z_j}, ~P^{(k)}_{X_j}$.  If a specific ket is affected by noise on site $j$ of copy $k$, it will be an eigenvector of $D _j ^{(k)}$ with zero eigenvalue.

The noise from a CNOT at site $j$ between copies $i$ and $k$ is
\begin{equation}
E_j^{(i),(k)} \equiv (1-p_2) + p_2 (P^{(i)}_{Z_j}  P^{(i)}_{X_j} )(P^{(k)}_{Z_j} P^{(k)}_{X_j} ).
\end{equation}
If a ket $\kket{\ab,\bb}$ is affected by any of these noise terms (that is, if the noise anticommutes with $\stab{\ab,\bb}$), it will be projected to zero and thus acquire a $(1-p_2)$ multiplier overall.

The noise from the first MCNOT is $E_{01} \equiv \prod_j E_j^{(0),(1)}$ and from the second MCNOT is $E_{02} \equiv \prod_j E_j^{(0),(2)}$. Clearly the overall multiplier is independent of the measurement outcomes, so the analysis for \Eref{eq:proj} still holds. The recursion relations are then similar in structure to \Eref{eq:gen_stab_rec}, except that coefficients dependent on $(1-p_2)$ are inserted before each term.

We illustrate this by calculating the recursion relations for $\avg{j}$.  If $j\in B$, there is no sum in \Eref{eq:gen_stab_rec}, and $\avg{j} \rightarrow E_j \avg{j}^{3}$.  The only noise terms that anticommute with  $\gen_j$ [and hence give factors of $(1-p_2)$] are those in $j \cup N_j$.  There are $2(d+1)$ of these (since there are two sets of noisy gates), so $\avg{j} \rightarrow (1-p_2)^{2(d+1)}\avg{j}^3$, which is \Eref{eq:3copyn:b}.

Now suppose $j \in A$. Let $\vec{j} = (0,\dots,0,j,0,\dots,0)$. Our sum is over $\aone, \atwo \in \{\vec{0},\vec{j} \}$, and $\bb = \vec{0}$. Since we are interested only in \avgj, our effective noise model is $[\X]_k \mapsto [\Z]_j \forall k \in N_j$ and $[\X]_j \mapsto [\I]$.  All other noise terms do not affect the state. Then
\begin{equation}\label{eq:effnoise}
E_{01} \mapsto \left[ (1-p_2) + p_2 P_Z^{(j,0)}P_Z^{(j,1)}\right]^{d+1}.
\end{equation}
A similar replacement holds for $E_{02}$. $E_{01}$ acts on terms $\kket{\vec{j}+\ab^{(1)}+\ab^{(2)},\vec{0}}\kket{\ab^{(1)},\vec{0}}$ and gives a factor of 1 iff $\vec{j}+\ab^{(1)}+\ab^{(2)}=\vec{0},\ab^{(1)}=\vec{0} \Rightarrow \vec{j}=\ab^{(2)},\ab^{(1)}=\vec{0}$, and a factor of $(1-p_2)^{d+1}$ otherwise.

Performing the MCNOT between $\zero$ and $\two$, the noise channel $E_{02}$ acts on the kets $\kket{\vec{j}+\ab^{(1)},\vec{0}}\kket{\ab^{(2)}}$, which gives a factor of 1 iff $\vec{j}+\ab^{(1)}=\vec{0},\ab^{(2)}=\vec{0}$ and $(1-p)^{d+1}$ otherwise.  Putting in each of the four cases $a^{(1)}_j,a^{(2)}_j \in \bin$ gives us \Eref{eq:3copyn:a}.

\subsection{Behavior of correlations}\label{sec:3copy:correlations}

If we take two qubits $j,k$ such that $\neigh(j) \cap \neigh(k) = \oslash$, then the noise terms on sites in $\neigh(k) \cup k$ do not affect terms involving $j$ and vice versa.  Hence the sum over terms in the recursion relation for $\avg{jk}$ will factor into $\avg{j}\avg{k}$.  If initially $\expec{\gen_j\gen_k} = \avg{j} \avg{k}$, then the three-copy protocol will not generate any new correlations between these regions.

\section{Uniqueness of the fixed point}\label{sec:fixed:points}
Here we show that the three-copy protocol has a unique fixed point for stabilizer elements $\expec{\gen_{\ab,\bb}}$ with weight $w=\vectornorm{\ab}+\vectornorm{\bb}\leq 2$. The recursion relations for stabilizer elements of weight $w>1$ [see \Eref{eq:gen_stab_rec}] depend only on stabilizer elements whose weight is at most $w$. Thus, we can use an inductive argument. If all the stabilizer elements of weight less than $w$ have reached a fixed point, they become constants and then the recursion relation for elements of weight $w$ will have the same form as those for weight one (i.e., they will depend only on stabilizer elements of weight $w$). First consider the case when $\vectornorm{\ab},\vectornorm{\bb}\leq 1$. For this case, the three-copy recursion relations \Eref{eq:gen_stab_rec} have the form
\begin{align*}
f(z) &= az + b z^3,\\
g(z) &= cz + d z^3,
\end{align*}
with $a,c > 0$ and $bd < 0$. The presence of noise does not change the form of the recursion relations, it only multiplies each term by a number between 0 and 1 (see Appendix~\ref{sec:recurrence:noisy}). Let $y = z^2$ and $x = d y + c$. Define
\begin{equation*}
  p(x) := f(g(z))/z - 1 = b x^4 - bc x^3 + adx - d.
\end{equation*}
The signature of $p(x)$ is
\begin{align*}
  p(x)&:-++-,\\
  p(-x)&:----.
\end{align*}
Then by Descartes' rule of signs \cite{Descartes1954}, $p(x)$ has at least two complex roots. Thus the recursion relation $f(g(z))=z$ has at most two positive fixed points. The recursion relation $g(f(z))=z$ can be analyzed identically. It was already argued in \Sref{sec:3copy:noisy} that this means that there is a unique attractive fixed point.

Now consider the case $\vectornorm{\ab} = 2$ and $\vectornorm{\bb}=0$. The recursion relations now have the form
\begin{align*}
f(z) &= az^3 + b z + c\\
g(z) &= d z^3.
\end{align*}
It is easily checked that $a$, $c$, and $d$ are positive. The sign of $b$ is harder to fix, but we note that for there to be a fixed point at all, $b$ must be negative. The case $f(g(z))=z$ is easily analyzed, as above, to show that there are at most two positive roots. Let $p(z)=g(f(z))$. To conclude the proof we need two technical results. ({\romannumeral 1}) If the smallest support  expectation value $\avg{a}$ has reached its fixed point value $\avg{a}_{\text{fp}}$ then the physically allowed values for $\avg{a+a'}$ form the interval $I = [2\avg{a}_{\text{fp}}-1,1]$. ({\romannumeral 2}) $f(z) \geq 0$ for all $z \in I$. Proof of ({\romannumeral 1}) (a) $z$ allowed $\Rightarrow z \in I$: $P= \frac{1 -\gen_a}{2}\frac{1 -\gen_{a'}}{2}$, with $a \neq a'$, is a projector, hence $\expec{P} \geq 0$. Thus $z = \expec{\gen_{a+a'}} \geq \avg{a} + \avg{a'} -1$ (*). Evaluate (*) at fixed point $\avg{a}_{\text{fp}}$. $z \leq 1$ is obvious. (b) $z \in I \Rightarrow z$ allowed: For an initial state of the protocol, interpolate between $\rho_1 = \avg{a}_{\text{fp}} \rho_{++} + (1-\avg{a}_{\text{fp}})/2\, (\rho_{+-}+\rho_{-+})$ and $\rho_2= \avg{a}_{\text{fp}} \rho_{++} + (1-\avg{a}_{\text{fp}}) \rho_{--}$. (The signs ``$\pm$'' refer to the eigenvalues of $\gen_a$ and $\gen_{a'}$, respectively.) Proof of ({\romannumeral 2}). Be $\avg{a}_{\text{fp}}, \avg{b}_{\text{fp}}  > 0$ and $z \in I$. Assume as an hypothesis $f(z) < 0$. Apply (*) to the state after application of P1, at the fixed point $\avg{a}_{\text{fp}}, \;\forall a \in \A$. Hence $0 \geq f(z) \geq 2 \avg{b}_{\text{fp}} -1$. (Under P1 the fixed point value $\avg{a}_{\text{fp}}$ for  $a\in \A$ is mapped to  $\avg{b}_{\text{fp}}$ for  $b\in \B$, assuming all vertices have the same degree.)  Thus, $\avg{b}_{\text{fp}} \leq 1/2$. But then $\avg{b}_{\text{fp}}=0$, which is a contradiction. Hence $f(z) \geq 0$.

Now, $p^{\prime\prime}(z) = g^{\prime\prime}(f(z))f^\prime(z)^2+g^\prime(f(z))f^{\prime\prime}(z)$ such that, with ({\romannumeral2}),  $p^{\prime\prime} \geq 0$ for all $z \in I$. Thus, $p(z)$ is   convex on $I$. With ({\romannumeral 1}), $I$ is a single interval such that $p(z)$ and $z$ intersect at most twice in $I$. At most one of these fixed points is attractive.

\section{The depolarizing operator}
\label{app:general}
In order to prove that the depolarizing operator \D defined in \Eref{eq:depolOp} commutes with the evolution operator $\R=  \Tr_{(1,2)} M \circ \Err \circ U$, we note that the protocol step P1 consists of a unitary part $U$, an error channel \Err comprising probabilistic Pauli errors, and a measurement $\Tr_{(1,2)} M$, where $M$ is a projector. $U$ consists of a
set of transversal CNOT-gates and acts on the stabilizer as
\begin{equation}
  \begin{array}{rcl}
    \gen^{(0)}_{{\vec{a}},{\vec{b}}} & \longrightarrow &
    \gen^{(0)}_{{\vec{a}},{\vec{b}}}  \gen^{(1)}_{{\vec{a}},{\vec{0}}}
    \gen^{(2)}_{{\vec{a}},{\vec{0}}},\\
    \gen^{(1)}_{{\vec{a}},{\vec{b}}} & \longrightarrow &
    \gen^{(1)}_{{\vec{a}},{\vec{b}}}  \gen^{(0)}_{{\vec{0}},{\vec{b}}},\\
    \gen^{(2)}_{{\vec{a}},{\vec{b}}} & \longrightarrow &
    \gen^{(2)}_{{\vec{a}},{\vec{b}}}  \gen^{(0)}_{{\vec{0}},{\vec{b}}}.
  \end{array}
\end{equation}
Now note that $\frac{[I]+[\gen_{\vec{0},\vec{b}}^{(1)} \gen_{\vec{0},\vec{b}}^{(0)}]}{2}
\frac{[I]+[\gen_{\vec{0},\vec{b}}^{(0)}]}{2}  \frac{[I]+[\gen_{\vec{0},\vec{b}}^{(2)} \gen_{\vec{0},\vec{b}}^{(0)}]}{2}=
\frac{[I]+[\gen_{\vec{0},\vec{b}}^{(0)}]}{2} \frac{[I]+[\gen_{\vec{0},\vec{b}}^{(1)}]}{2}
\frac{[I]+[\gen_{\vec{0},\vec{b}}^{(2)}]}{2}$ etc., such that
\begin{equation}
  \label{eq:comm2}
  U \circ \D^{(0)} \D^{(1)} \D^{(2)} = \D^{(0)}
  \D^{(1)} \D^{(2)} \circ U.
\end{equation}
The operations ${\D}^{(0)}  {\D}^{(1)}  {\D}^{(2)}$
and ${\Err}$ commute because both are linear combinations of Pauli superoperators,
\begin{equation}
  \label{eq:commDE}
  {\Err} \circ {\D}^{(0)}  {\D}^{(1)}  {\D}^{(2)}  =
  {\D}^{(0)} {\D}^{(1)} {\D}^{(2)}  \circ {\Err}.
\end{equation}
The measurements comprising $\Tr_{(1,2)} M$ are of stabilizer operators
$\gen_{\vec{0},\vec{b}}^{(1)}$, $\gen_{\vec{0},\vec{b}}^{(2)}$ on the states $\one$,
$\two$, respectively. They are performed via one-qubit measurements
and classical post-processing. $\gen_{\vec{0},\vec{b}}^{(1)}$, $\gen_{\vec{0},\vec{b}}^{(2)}$
commute with the Kraus operators in \Eref{eq:depolOp}, such that
\begin{align}
  \label{eq:comm3}\notag
  \Tr_{(1,2)} M \circ  \D^{(0)} \D^{(1)} \D^{(2)} &=  \Tr_{(1,2)} \D^{(0)} \D^{(1)} \D^{(2)}  \circ M \\
  &=  \D^{(0)}  \circ  \Tr_{(1,2)}  M.
\end{align}
Eqs.~(\ref{eq:comm2}), (\ref{eq:commDE}), and (\ref{eq:comm3}) yield \Eref{eq:comm}
\section{Creation of a bi-colorable graph state}\label{sec:creation}

\begin{figure}
\includegraphics{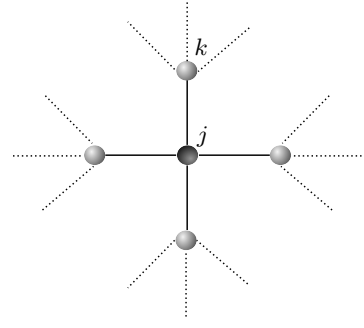}
\caption{\label{fig:creation}Creation of a degree ($d=4$) bicolorable graph state. The figure will have the same local structure for other degrees and topologies as long as its edges are $d$ colorable and its vertices are bicolorable}
\end{figure}

Here we discuss the noise structure of a bicolorable graph that is created using noisy CPHASE gates. The noisy gates are modeled as the ideal gate followed by two-qubit depolarizing noise as defined in \Eref{eq:depol}. The graph state is created by performing CPHASE gates between qubits in the $\ket{+}$ state. The noise structure of the final state depends on the temporal ordering of these gates. If we assume that the underlying graph has constant degree $d$ and that its edges are $d$ colorable, then the $N$-qubit graph state can be created in $d$ time steps with $Nd$ CPHASE gates. At each time step all the gates corresponding to edges of a particular color are performed. Thus, at every time step $t\in\{1,\dots,d\}$, each qubit is affected by an error channel of the form of \Eref{eq:depol}.

We are interested in the value of \avgj, so we focus on the neighborhood of qubits $j$ in the larger graph. Since the graph is bicolorable, it contains no three cycles and one can draw a diagram of the form of \Fref{fig:creation}. The gates are represented by both solid as well as dashed lines. The noise channels corresponding to the solid lines each contribute an effective error $\T_{\text{eff}}$ as defined in \Eref{eq:flipNoise} to qubit $j$. Now consider the qubit $k$ which is a neighbor of the central qubit $j$. Each dashed line also contributes an effective error $\T_{\text{eff}}$ to qubit $j$, but only if the CPHASE gate corresponding to the solid line between $k$ and $j$ was performed in a previous timestep. This is because $\Z_k$ errors commute with $\gen_j$ and $\X_k$ errors would be propagated by the CPHASE to $\X_k\Z_j$ errors, which also commute with $\gen_j$. Thus there are a total of $\frac{d(d-1)}{2} + d = \frac{d(d+1)}{2}$ noise channels affecting the qubit $j$. This gives
\begin{equation}\label{eq:creation}
         \avgj = (1-p_2)^{\frac{d(d+1)}{2}}.
\end{equation}

\end{document}